# Efficacy of Temporal Interference Electrical Stimulation for Spinal Cord Injury Rehabilitation: A Case Series

Ruidong Cheng [a,b], Yuling Shao [a], Xi Li [c], Li Zhang [a,b], Zehao Sheng [a], Chenyang Li [d], Xu Xie [d], Huilin Mou [d], Weidong Chen [d], Shaomin Zhang [d,*], Yuchen Xu [d,e,*], MinminWang [d,f,*]

[a] Center for Rehabilitation Medicine, Rehabilitation & Sports Medicine Research Institute of Zhejiang Province, Department of Rehabilitation Medicine, Zhejiang Provincial People' s Hospital (Affiliated People's Hospital, Hangzhou Medical College), Hangzhou, China.
[b] School of rehabilitation, Hangzhou Medical College, Hangzhou, China.
[c] Graduate School, Hangzhou Normal University, Hangzhou, China.
[d] Qiushi Academy for Advanced Studies, Zhejiang University, Hangzhou, China.
[e] Center of Excellence in Biomedical Research on Advanced Integrated-on-Chips Neurotechnologies (CenBRAIN Neurotech), School of Engineering, Westlake University, Hangzhou, China
[f] Westlake Institute for Optoelectronics, Westlake University, HangZhou, China.

*For Correspondence:

Minmin Wang, minminwang@zju.edu.cn, Qiushi Academy for Advanced Studies, Zhejiang University, Hangzhou 310027, China.

**Abstract**

*Background:* Spinal cord injury (SCI) is a debilitating condition that often results in significant motor and sensory deficits, impacting the quality of life. Current rehabilitation methods, including physical therapy and electrical stimulation, offer variable outcomes and often require invasive procedures. Temporal interference (TI) stimulation has emerged as a novel, non-invasive neuromodulation technique capable of targeting deep neural structures with precision, providing a promising alternative for SCI rehabilitation. This study explores the efficacy of TI stimulation as a non-invasive approach for improving motor and sensory function in patients with incomplete SCI.

*Methods:* Three male patients with incomplete cervical SCI (AIS D) participated in a two-week intervention consisting of 14 sessions of TI stimulation targeting their injury sites. TI stimulation was delivered using frequencies of 1000 Hz and 1040 Hz, with assessments conducted pre- and post-intervention, including motor and sensory evaluations, functional scales, and imaging studies.

*Results:* All participants demonstrated significant improvements in neurological function, motor strength, sensory perception, and functional independence. Neurological levels of injury shifted upward in all cases, with one patient improving from C5 to C7. Graded Redefined Assessment of Strength, Sensibility and Prehension (GRASSP) results shows additional strength, prehension and sensory outcomes obtained for the arm and hand functions of participants. Motor scores (UEMS and LEMS) increased, sensory scores for light touch and pin prick improved, and functional assessments, such as the Berg Balance Scale (BBS) and Barthel Index (BI), showed marked gains. Pain scores also decreased in two participants, highlighting additional therapeutic benefits.

*Conclusions:* TI stimulation facilitated meaningful recovery in motor, sensory, and functional domains, suggesting its potential as a novel, effective, and non-invasive rehabilitation strategy for incomplete SCI.

# 1. Introduction

Spinal cord injury (SCI) is a devastating condition that often leads to neurological deficits, severely affecting the quality of life of those who are affected [1]. The impact of SCI is profound, as it can result in complete or partial paralysis, loss of sensation, and disruption of autonomic functions [2, 3]. Despite significant advancements in medical and surgical treatments, the recovery of motor and sensory functions remains limited [4]. The challenge of restoring neurological function in these patients underscores the need for innovative rehabilitation strategies. Rehabilitation plays a crucial role in promoting functional recovery, enhancing mobility, and improving the overall well-being of individuals with SCI [5]. Traditional rehabilitation methods, such as physical therapy, occupational therapy, and pharmacological interventions, are aimed at stimulating neural plasticity, encouraging functional recovery, and improving the quality of life [6, 7]. However, the effectiveness of these methods is variable, and they often fail to provide significant improvements in patients with severe injuries or incomplete SCI.

One of the most promising techniques in SCI rehabilitation is spinal cord stimulation (SCS), which involves the application of electrical pulses to activate spinal neurons and stimulate motor function recovery [8, 9]. SCS has been widely studied and used to treat chronic pain and motor dysfunction in SCI patients, with varying degrees of success [10, 11]. The technique shows potential for enhancing functional recovery by modulating the activity of spinal circuits. However, the effectiveness of electrical stimulation is often dependent on several factors, including the severity and level of the injury, as well as the individual's response to treatment [12, 13]. Additionally, the invasive nature of traditional spinal cord stimulation, which requires surgical implantation of electrodes, limits its widespread clinical application and patient accessibility [14].

In response to these limitations, temporal interference (TI) electrical stimulation has emerged as a promising non-invasive alternative for neuromodulation [15]. TIS involves applying two electrical currents with slightly different frequencies to the targeted area, generating an interference pattern that delivers a low-frequency stimulation to deeper tissues [16, 17]. The resulting envelope modulation enhances the regulation of neural activity, offering a potential solution for neuromodulation without the need for invasive procedures [18, 19]. This non-invasive nature of TI stimulation makes it an appealing option for rehabilitation, particularly for patients with SCI, where non-invasive approaches are preferred due to the risks associated with invasive treatments.

TI electrical stimulation has already been explored in preliminary studies for various clinical applications, such as movement disorder, depression, and neural regeneration [20-23]. Its ability to modulate neural circuits makes it a potentially valuable tool for SCI rehabilitation, particularly for individuals with incomplete spinal cord injuries. In incomplete SCI, some neural pathways remain intact, and TI stimulation may enhance the remaining neural activity, promoting neural plasticity and functional recovery [24-28]. This modulation of neural circuits may provide significant benefits in patients with partial preservation of spinal cord function, offering a promising alternative to traditional therapies that may not yield satisfactory results in clinical practice.

In this study, we aim to evaluate the efficacy of TIS in the rehabilitation of patients with incomplete spinal cord injury.

## 2. Methods and Materials

### 2.1 Participants

The trial included adult participants aged 18–80 years who had sustained a traumatic, non-progressive cervical (C4–C8) SCI between 1 month and 6 months. Participants with American Spinal Injury Association (ASIA) Impairment Scale (AIS) classification B, C or D were considered for enrollment. The participants remain on stable medications throughout the study. All participants were capable of providing written informed consent.

Inclusion criteria were as follow: (1) Age 18-80 (women or men); (2) Incomplete SCI graded as AIS B, C & D; (3) Level of lesion: Non-progressive cervical SCI from C4–C8 inclusive; (4) Onset of SCI between 1 month and 6 months; (5) ASIA assessment has not changed in the past week; (6) Indicated for rehabilitation training procedures by the participant's treating physician, occupational therapist or physical therapist; (7) Remain on stable medications throughout the study; (8) Agree to comply with all conditions of the study and to attend all required study training and rehabilitation assessments; (9) Must sign Informed Consent prior to any study related procedures.

Exclusion criteria were as follow: (1) Limitation of motor function based on neurologic diseases such as stroke, multiple sclerosis and traumatic brain injury; (2) Has any unstable or significant medical conditions; (3) Epilepsy; (4) Contraindications of electrical stimulation therapy.

This case series included three participants with spinal cord injuries:

Participant 1 (P1): A 34-year-old male who sustained SCI due to a vehicular accident 3.5 months prior to the study. The injury was classified as AIS D with a neurological level of injury at C5.

Participant 2 (P2): A 39-year-old male with SCI caused by an intraspinal hematoma (spinal cord vascular malformation) 5 months prior to the study. The injury was classified as AIS D with a neurological level of injury at C4.

Participant 3 (P3): A 70-year-old male who sustained SCI due to a vehicular accident 3 months prior to the study. The injury was classified as AIS D with a neurological level of injury at C4.

Based on assessments conducted one week before and at the start of the TI stimulation, no changes were observed in the severity of spinal cord injury among the three participants. Ethical approval for this study was granted by the Ethics Committee of Zhejiang Provincial People's Hospital (No. KY2024103). All participants provided informed consent, agreeing to the use of their data in future publications and presentations.

**2.2 Study Design and Assessments**

The participants enrolled in the clinical trial underwent a standardized rehabilitation program with the addition of TI stimulation. The study involved 14 sessions of TI stimulation for each participant. TI stimulation was applied to the spinal cord injury site using frequencies of 1000 Hz and 1040 Hz, with each session lasting 20 minutes and a maximum stimulation current of 4 mA.

Participants were assessed using various a set of clinical scales at baseline (pre-intervention) and at the end of the experimental session (during two weeks) after stimulation therapy. The assessments included:

Motor and functional scales: Graded Redefined Assessment of Strength, Sensation and Prehension (GRASSP), Upper Extremity Fugl-Meyer (UEFM), Lower Extremity Fugl-Meyer (LEFM), American Spinal Injury Association (ASIA) impairment scale, Berg Balance Scale (BBS), and Barthel Index (BI).

Pain assessments: Short-Form McGill Pain Questionnaire (SF-MPQ), urinary diaries.

Imaging and neurophysiological evaluations: Electromyography (EMG), magnetic resonance imaging, and computed tomography.

**2.3 Treatment Protocol of Temporal Interference Stimulation**

TI was used to target the spinal cord injury site at the cervical region with precision. The protocol involved five steps:

Step 1: Acquisition of Individual Imaging Data

High-resolution neuroimaging data were collected to guide precise targeting. Imaging included MRI and CT scans:

MRI: Performed on a 1.5T or higher field-strength scanner to capture soft tissue and neural structures;

CT: Acquired using a 64-slice multidetector scanner to visualize vertebral and spinal canal structures.

Step 2: Construction of Individual Cervical-Thoracic Models

Using the acquired imaging data, individualized three-dimensional anatomical models of the cervical region were constructed. Segmentation of soft tissues in MRI data was performed using a convolutional neural network-based tool (nnUNet). Bone structures and spinal canal from CT data were segmented using template matching methods. 3D Slicer software was used to generate high-resolution models and validate anatomical landmarks (e.g., vertebral boundaries) for accuracy. Adjustments were made as necessary. The final models included various tissue types (e.g., subcutaneous muscle, spine, spinal cord) with distinct dielectric properties.

Step 3: Candidate Electrode Placement and Meshing

Electrode placement was optimized to maximize stimulation coverage of the target region while minimizing off-target effects: Electrode Placement: Candidate electrodes were symmetrically positioned around the target region on the skin surface, with 1 cm increments in both lateral and vertical directions. Placement was aligned with neural root projections. Meshing: A tetrahedral mesh was applied to the model. The target region used a high-density mesh, while coarser meshes were used for peripheral regions to reduce computational load.

Step 4: Leadfield Calculation for Candidate Electrode Arrays

Leadfield calculations were performed to simulate electric field distributions generated by TI stimulation:

Simulation: Three-dimensional electric field simulations were conducted using COMSOL Multiphysics with input parameters of 1 mA current.

Methodology: Finite element method (FEM) was used to compute electric field distributions.

Step 5: Optimization of Stimulation Parameters

The efficacy of tTIS depends on maximizing the envelope electric field ($|\vec{E}^{MAX}(\vec{r})|$) in the target region while suppressing its effects in non-target areas. The envelope electric field ($|\vec{E}^{MAX}(\vec{r})|$ was computed using the following formula:

$$|\vec{E}^{MAX}(\vec{r})| = \begin{cases} 2|\vec{E}_2(\vec{r})|, & if\ |\vec{E}_2(\vec{r})| < |\vec{E}_1(\vec{r})|\cos(a) \\ \frac{2|\vec{E}_2(\vec{r}) \times (\vec{E}_1(\vec{r}) - \vec{E}_2(\vec{r}))|}{|\vec{E}_1(\vec{r}) - \vec{E}_2(\vec{r})|}, & otherwise \end{cases}$$

Where $|\vec{E}_1(\vec{r})|$ and $|\vec{E}_2(\vec{r})|$ represent the maximum amplitudes of the two electric fields generated by two electrode pairs: $C_{11} - C_{12}$ (current $I_1$) and $C_{21} - C_{22}$ (current $I_2$). The respective fields were calculated using the following equations:

$$\vec{E_1} = I_1 * (C_{11} * Leadfield - C_{12} * Leadfield\ )$$
$$\vec{E_2} = I_2 * (C_{21} * Leadfield - C_{22} * Leadfield\ )$$

To achieve envelope-specific stimulation, customized electrode configurations were developed for each participant using individual magnetic resonance imaging (MRI) or CT data.

Once the optimal tTIS parameters (electrode positions and stimulation currents) were determined, the stimulation was administered using a tTIS device and Ag/AgCl electrodes with hydrogel were used to ensure good conductivity and minimize skin impedance.

## 3. Results

### 3.1 Function Assessment

As shown in Table 1, the outcomes of the two-week TI stimulation intervention demonstrated significant improvements in motor, sensory, and functional assessments across all three participants with incomplete spinal cord injuries. Notably, all participants experienced an upward shift in their neurological levels of injury, indicating potential recovery of spinal cord function. Participant 1 improved from C5 to C7, Participant 2 from C4 to C5, and Participant 3 from C4 to C6.

The Graded Redefined Assessment of Strength, Sensibility and Prehension (GRASSP) results report the improvements across individual outcomes that were measured 2 weeks for additional strength, function and sensory outcomes obtained for the participants who completed the course of TI Therapy.

In Participant 1, GRASSP total (right/left) score improved from 97/96 to 99/108, GRASSP-Strength (right/left) score (37/40 to 39/44), Prehension ability (right/left) (11/11 to 11/12), GRASSP-Prehension Performance score (right/left) (27/26 to 30/30), Sensibility (Dorsal + Palmar) (left) (19 to 22), only Sensibility (Dorsal + Palmar) score (right) was reduced (22 to 19), taking into account the effect of hypertonia in the patient's left hand.

In Participant 2, GRASSP total (right/left) score improved from 83/106 to 95/111, GRASSP-Strength (right/left) score (36/48 to 40/50), Sensibility (Dorsal + Palmar) (right/left) (10/16 to 16/19), Prehension ability (right/left) (11/12 to 12/12), GRASSP-Prehension Performance (right/left) score (26/30 to 27/30).

In Participant 3, GRASSP total (right/left) score improved from 80/77 to 88/87, GRASSP-Strength (right/left) score (31/32 to 34/36), Sensibility (Dorsal + Palmar) (right/left) (13/9 to 15/12), GRASSP-Prehension Performance (right/left) score (24/24 to 27/27). There was no change in prehension ability (right/left) (12/12 to 12/12).

Motor function showed consistent gains in both upper and lower extremities. The total Upper Extremity Motor Scores (UEMS) increased in all patients, with notable improvements in Participant 2 from 40 to 46. Similarly, Lower Extremity Motor Scores (LEMS) demonstrated significant recovery, particularly in Participant 1, whose score increased from 36 to 43. These gains suggest enhanced motor control and strength following TI stimulation.

Sensory assessments revealed marked enhancements in light touch and pin prick sensitivity. For instance, Participant 1's light touch sensory score increased from 64 to 80, while Patient 2 showed improvements in pin prick sensitivity from 83 to 87. These findings indicate that TI stimulation may contribute to restoring sensory pathways and improving overall perception in patients with incomplete SCI.

Functional assessments highlighted considerable progress in balance, independence, and pain reduction. The Berg Balance Scale (BBS) scores improved across all participants, with the most striking increase observed in Participant 3, whose score rose from 4 to 36. Similarly, the Barthel Index (BI) scores, reflecting daily functional independence, improved in all patients, particularly in Participant 2, who achieved an increase from 50 to 70. Additionally, pain scores measured by the SF-MPQ decreased in two patients, further supporting the therapeutic benefits of TI stimulation.

Overall, the results demonstrate that TI stimulation can facilitate meaningful recovery in motor, sensory, and functional domains for individuals with incomplete SCI within a relatively short intervention period. These findings highlight the potential of TI as a promising and effective non-invasive rehabilitation modality.

TABLE 1 Demographic and clinical characteristics of the participants

| Participant | P1 | | P2 | | P3 | |
|---|---|---|---|---|---|---|
| Sex | Male | | Male | | Male | |
| Age (years) | 34 | | 39 | | 70 | |
| Time of injury (months) | 3.5 | | 5 | | 3 | |
| Cause of injury | Vehicular | | Intraspinal hematoma（spinal cord vascular malformation） | | Vehicular | |
| Assessments at study enrollment (Pre) And after rehabilitation period (Post) | Pre | Post | Pre | Post | Pre | Post |
| American Spinal Injury Association Impairment Scale (AIS) | D | D | D | D | D | D |
| Neurological level of injury | C5 | C7 | C4 | C5 | C4 | C6 |
| Upper Extremity Motor Scores (right/left) | 19/20 | 19/21 | 18/22 | 21/25 | 15/16 | 17/17 |
| Total | 39 | 40 | 40 | 46 | 31 | 34 |
| Lower Extremity Motor Scores (right/left) | 17/19 | 20/23 | 8/21 | 11/21 | 21/20 | 22/22 |
| Total | 36 | 43 | 29 | 32 | 41 | 44 |
| Light Touch Sensory Scores (right/left): | 32/32 | 40/40 | 30/29 | 32/32 | 31/31 | 35/35 |
| Total | 64 | 80 | 59 | 64 | 62 | 70 |
| Pin Prick Sensory Scores (right/left) | 26/32 | 37/39 | 51/32 | 53/34 | 32/31 | 34/33 |
| Total | 58 | 76 | 83 | 87 | 63 | 67 |
| **Graded Redefined Assessment of Strength, Sensation and Prehension （GRASSP）** | | | | | | |
| Strength (right/left) | 37/40 | 39/44 | 36/48 | 40/50 | 31/32 | 34/36 |
| Sensibility（Dorsal + Palmar） (right/left) | 22/19 | 19/22 | 10/16 | 16/19 | 13/9 | 15/12 |
| Prehension ability (right/left) | 11/11 | 11/12 | 11/12 | 12/12 | 12/12 | 12/12 |
| Prehension performance (right/left) | 27/26 | 30/30 | 26/30 | 27/30 | 24/24 | 27/27 |
| Total (right/left) | 97/96 | 99/108 | 83/106 | 95/111 | 80/77 | 88/87 |
| Fugl-Meyer upper extremity score (UEFM) | 65/64 | 65/66 | 62/64 | 63/66 | 53/58 | 58/58 |
| Fugl-Meyer assessment lower extremity score (LEFM) | 28/28 | 33/34 | 26/30 | 28/30 | 13/13 | 30/29 |
| Berg Balance Scale (BBS) | 44 | 50 | 36 | 45 | 4 | 36 |
| Barthel Index (BI) | 60 | 70 | 50 | 70 | 25 | 45 |
| SF-MPQ | 10 | 7 | 5 | 4 | 4 | 4 |

## 3.2 Adverse Effects

Adverse events were minor and transient. No severe or lasting adverse events were reported throughout the study, indicating the safety and tolerability of the TI intervention.

## 4. Discussion

This study presents a novel approach to temporal interference stimulation for treating spinal cord injury, specifically targeting cervical spinal cord regions in patients with incomplete injuries. By conducting interventions on three participants, our results demonstrated noticeable functional recovery after only two weeks of TI stimulation, offering new insights into non-invasive neuromodulation therapies for SCI rehabilitation.

The results underscore the potential of TI stimulation to facilitate functional recovery in patients with incomplete SCI. As a novel neuromodulation technique, TI leverages the intersection of high-frequency electrical fields to achieve deep, focused stimulation without activating superficial tissues. This capability addresses the primary challenge of many traditional neuromodulation methods—non-specific stimulation of peripheral regions. Our study also contributes to the growing body of evidence supporting individualized treatment protocols in neurorehabilitation. By combining advanced imaging, computational modeling, and tailored electrode configurations, we ensured precise targeting of spared neural pathways. This personalized approach could serve as a cornerstone for future therapies aimed at optimizing neuroplasticity and recovery. The individualized modeling and optimization process, while effective, required considerable computational resources and expertise. Efforts to streamline these processes will be critical for widespread clinical adoption. Future studies should aim to validate these findings, refine stimulation protocols, and explore broader clinical applications to fully realize TI's potential as a transformative therapy for neurological disorders.

Numerous studies have reported the efficacy of electrical stimulation in promoting recovery in SCI patients. For example, epidural spinal cord stimulation (eSCS) has demonstrated remarkable results in reactivating motor function in individuals with severe SCI [29]. However, eSCS typically requires invasive surgical implantation and carries risks such as infection or hardware-related complications [30, 31]. Compared to eSCS, TI offers a non-invasive alternative with potentially comparable benefits, making it a safer and more accessible option for broader patient populations.

Similarly, transcutaneous electrical stimulation (TES) has been used to enhance motor function in SCI patients by stimulating spinal circuits via skin electrodes [32, 33]. While TES has achieved some success, its inability to stimulate deep spinal cord regions restricts its therapeutic potential. TI, with its capacity for deep and focused stimulation, bridges this gap by accessing structures beyond the reach of TES.

Future research should explore these mechanisms in greater depth, incorporating biomarkers such as neuroimaging, electrophysiological recordings, and molecular assays. Additionally, studies should investigate how TI interacts with other rehabilitation strategies, such as physical therapy, functional electrical stimulation, and pharmacological agents, to create synergistic effects.

While our results are promising, several limitations should be addressed. The current study involved only three participants, limiting the statistical power and generalizability of our findings. Future research should include larger, diverse cohorts to validate these results; Longer trials with extended follow-ups are necessary to evaluate the durability of TI-induced benefits; All participants had incomplete injuries classified as AIS D, indicating preserved motor function. Future studies should include patients with more severe injuries, such as AIS A or B, to evaluate the applicability in cases with minimal neural preservation.

## 5. Conclusion

In conclusion, the functional recovery observed in the three participants of this study demonstrates the potential of temporal interference stimulation as an effective and non-invasive approach for spinal cord rehabilitation.

## Conflict of Interest

All authors declare no competing interests.

## Data availability

Data involved in this study are available upon reasonable request.

## Acknowledgments

We would like to extend our heartfelt gratitude to all the patients who participated in this clinical study and their families for their invaluable support.

## Source of funding

Research supported by the China Brain Project (2021ZD0200401), the "Pioneer" and "Leading Goose" R&D Program of Zhejiang (2023C03081, 2023C03026).

## References

[1]    Barbiellini Amidei C, Salmaso L, Bellio S, Saia M. Epidemiology of traumatic spinal cord injury: a large population-based study. Spinal Cord 2022;60(9):812-9.
[2]    Lu Y, Shang Z, Zhang W, Pang M, Hu X, Dai Y, et al. Global incidence and characteristics of spinal cord injury since 2000–2021: a systematic review and meta-analysis. BMC Med 2024;22(1):285.

[3]     Anjum A, Yazid MDi, Fauzi Daud M, Idris J, Ng AMH, Selvi Naicker A, et al. Spinal Cord Injury: Pathophysiology, Multimolecular Interactions, and Underlying Recovery Mechanisms. International Journal of Molecular Sciences 2020;21(20):7533.
[4]     Karsy M, Hawryluk G. Modern Medical Management of Spinal Cord Injury. Curr Neurol Neurosci Rep 2019;19(9):65.
[5]     Duan R, Qu M, Yuan Y, Lin M, Liu T, Huang W, et al. Clinical Benefit of Rehabilitation Training in Spinal Cord Injury: A Systematic Review and Meta-Analysis. Spine 2021;46(6):E398-E410.
[6]     Musselman KE, Shah M, Zariffa J. Rehabilitation technologies and interventions for individuals with spinal cord injury: translational potential of current trends. J Neuroeng Rehabil 2018;15(1):40.
[7]     Venkatesh K, Ghosh SK, Mullick M, Manivasagam G, Sen D. Spinal cord injury: pathophysiology, treatment strategies, associated challenges, and future implications. Cell Tissue Res 2019;377(2):125-51.
[8]     Anderson MA, Squair JW, Gautier M, Hutson TH, Kathe C, Barraud Q, et al. Natural and targeted circuit reorganization after spinal cord injury. Nat Neurosci 2022;25(12):1584-96.
[9]     Sdrulla AD, Guan Y, Raja SN. Spinal Cord Stimulation: Clinical Efficacy and Potential Mechanisms. Pain Practice 2018;18(8):1048-67.
[10]    Lorach H, Galvez A, Spagnolo V, Martel F, Karakas S, Intering N, et al. Walking naturally after spinal cord injury using a brain–spine interface. Nature 2023;618(7963):126-33.
[11]    Moritz C, Field-Fote EC, Tefertiller C, van Nes I, Trumbower R, Kalsi-Ryan S, et al. Non-invasive spinal cord electrical stimulation for arm and hand function in chronic tetraplegia: a safety and efficacy trial. Nat Med 2024;30(5):1276-83.
[12]    James ND, McMahon SB, Field-Fote EC, Bradbury EJ. Neuromodulation in the restoration of function after spinal cord injury. The Lancet Neurology 2018;17(10):905-17.
[13]    Bekhet AH, Bochkezanian V, Saab IM, Gorgey AS. The Effects of Electrical Stimulation Parameters in Managing Spasticity After Spinal Cord Injury: A Systematic Review. Am J Phys Med Rehabil 2019;98(6):484-99.
[14]    Kumru H, Flores A, Rodríguez-Cañón M, Soriano I, García L, Vidal-Samsó J. Non-invasive brain and spinal cord stimulation for motor and functional recovery after a spinal cord injury. RN 2020;70(12):461-77.
[15]    Grossman N, Bono D, Dedic N, Kodandaramaiah SB, Rudenko A, Suk H-J, et al. Noninvasive Deep Brain Stimulation via Temporally Interfering Electric Fields. Cell 2017;169(6):1029-41.e16.
[16]    Violante IR, Alania K, Cassarà AM, Neufeld E, Acerbo E, Carron R, et al. Non-invasive temporal interference electrical stimulation of the human hippocampus. Nat Neurosci 2023;26(11):1994-2004.
[17]    Liu R, Zhu G, Wu Z, Gan Y, Zhang J, Liu J, et al. Temporal interference stimulation targets deep primate brain. NeuroImage 2024;291:120581.
[18]    Guo W, He Y, Zhang W, Sun Y, Wang J, Liu S, et al. A novel non-invasive brain stimulation technique: "Temporally interfering electrical stimulation". Front Neurosci 2023;17.
[19]    Hummel FC, Wessel MJ. Non-invasive deep brain stimulation: interventional targeting of deep brain areas in neurological disorders. Nature Reviews Neurology 2024;20(8):451-2.

[20] Alania K, Borella J, Violante I, Martin E, Rhodes E, Butler CR, et al. Non-Invasive Temporal Interference Hippocampal Stimulation in Early Alzheimer's Disease. Alzheimer's & Dementia 2023;19(S21):e077433.
[21] Missey F, Ejneby MS, Ngom I, Donahue MJ, Trajlinek J, Acerbo E, et al. Obstructive sleep apnea improves with non-invasive hypoglossal nerve stimulation using temporal interference. Bioelectronic Medicine 2023;9(1):18.
[22] Demchenko I, Rampersad S, Datta A, Horn A, Churchill NW, Kennedy SH, et al. Target engagement of the subgenual anterior cingulate cortex with transcranial temporal interference stimulation in major depressive disorder: a protocol for a randomized sham-controlled trial. Front Neurosci 2024;18.
[23] Yang C, Xu Y, Du Y, Shen X, Li T, Chen N, et al. Transcranial temporal interference subthalamic stimulation for treating motor symptoms in Parkinson's disease: A pilot study. Brain Stimulation: Basic, Translational, and Clinical Research in Neuromodulation 2024;17(6):1250-2.
[24] Xin Z, Abe Y, Kuwahata A, Tanaka KF, Sekino M. Brain Response to Interferential Current Compared with Alternating Current Stimulation. Brain Sciences 2023;13(9):1317.
[25] Caldas-Martinez S, Goswami C, Forssell M, Cao J, Barth AL, Grover P. Cell-specific effects of temporal interference stimulation on cortical function. Communications Biology 2024;7(1):1076.
[26] Lin H-C, Wu Y-H, Ker M-D. Modulation of Local Field Potentials in the Deep Brain of Minipigs Through Transcranial Temporal Interference Stimulation. Neuromodulation: Technology at the Neural Interface 2024.
[27] Mojiri Z, Rouhani E, Akhavan A, Jokar Z, Alaei H. Non-invasive temporal interference brain stimulation reduces preference on morphine-induced conditioned place preference in rats. Sci Rep 2024;14(1):21040.
[28] Vieira PG, Krause MR, Pack CC. Temporal interference stimulation disrupts spike timing in the primate brain. Nature Communications 2024;15(1):4558.
[29] McHugh C, Taylor C, Mockler D, Fleming N. Epidural spinal cord stimulation for motor recovery in spinal cord injury: A systematic review. NeuroRehabilitation 2021;49:1-22.
[30] Mansour NM, Peña Pino I, Freeman D, Carrabre K, Venkatesh S, Darrow D, et al. Advances in Epidural Spinal Cord Stimulation to Restore Function after Spinal Cord Injury: History and Systematic Review. J Neurotrauma 2022;39(15-16):1015-29.
[31] Chalif JI, Chavarro VS, Mensah E, Johnston B, Fields DP, Chalif EJ, et al. Epidural Spinal Cord Stimulation for Spinal Cord Injury in Humans: A Systematic Review. Journal of Clinical Medicine 2024;13(4):1090.
[32] Tefertiller C, Rozwod M, VandeGriend E, Bartelt P, Sevigny M, Smith AC. Transcutaneous Electrical Spinal Cord Stimulation to Promote Recovery in Chronic Spinal Cord Injury. Frontiers in Rehabilitation Sciences 2022;2.
[33] Mangold S, Keller T, Curt A, Dietz V. Transcutaneous functional electrical stimulation for grasping in subjects with cervical spinal cord injury. Spinal Cord 2005;43(1):1-13.